\documentclass{mn2e}
\usepackage{psfig}
\usepackage[authoryear]{natbib}

\newcommand\ros{{\it ROSAT}}

\newcommand\xmm{{\it XMM-Newton}}

\newcommand\rxja{RXJ005734}
\newcommand\rxjb{RXJ094144}
\newcommand\rxjc{RXJ121803}
\newcommand\rxjd{RXJ124913}
\newcommand\rxje{RXJ163308}

\newcommand\nh{N_{H}}

\title{The nature of X-ray absorbed QSOs}
\author[Page, et al.]{
M.J. Page\(^{1}\),
F.J. Carrera\(^{2}\),
J.A. Stevens\(^{3}\),
J. Ebrero\(^{4}\),
A.J. Blustin\(^{1,5}\)
\\
\(^{1}\)Mullard Space Science Laboratory, University College London,
Holmbury St Mary, Dorking, Surrey RH5 6NT, UK.\\
\(^{2}\)Instituto de F\'\i sica de Cantabria 
(CSIC--Universidad de Cantabria), 39005
Santander, Spain.\\
\(^{3}\)Centre for Astrophysics Research, University of Hertfordshire, 
College Lane, Herts, AL10 9AB, UK\\
\(^{4}\)SRON National Institute for Space Research, Sorbonnelaan 2, 3584 CA Utrecht, The Netherlands\\
\(^{5}\)Institute of Astronomy, University of Cambridge, Madingley Road, Cambridge CB3 0HA, UK\\
}

\date{}

\begin{document}
\maketitle

\begin{abstract}
There exists a significant population of broad line, $z\sim 2$ QSOs which have
heavily absorbed X-ray spectra. Follow up observations in the
submillimetre show that these QSOs are embedded in ultraluminous
starburst galaxies, unlike most unabsorbed QSOs at the same redshifts
and luminosities. Here we present X-ray spectra from \xmm\  
for a sample of 5 such X-ray absorbed QSOs that have been detected at
submillimetre wavelengths. We also present spectra in the restframe
ultraviolet from ground based telescopes. All 5 QSOs are found to
exhibit strong C\,IV absorption lines in their ultraviolet spectra
with equivalent width $> 5$\AA. The X-ray spectra are inconsistent
with the hypothesis that these objects show normal QSO continua
absorbed by low-ionization gas. Instead, the spectra can be modelled
successfully with ionized absorbers, or with cold absorbers if they
posess unusually flat X-ray continuum shapes and unusual optical to
X-ray spectral energy distributions. We show that the ionized absorber 
model provides the simplest, most self-consistent
explanation for their observed properties. We estimate that the
fraction of radiated power that is converted into kinetic luminosity
of the outflowing winds is typically $\sim$ 4 per cent, in agreement
with recent estimates for the kinetic feedback from QSOs required to
produce the $M - \sigma$ relation, and consistent with the hypothesis
that the X-ray absorbed QSOs represent the transition phase between
obscured accretion and the luminous QSO phase in the evolution of
massive galaxies.
\end{abstract}

\section{Introduction}

The prevalence of black holes in present-day galaxy bulges, and the
proportionality between black hole and spheroid mass 
\citep{magorrian98, merritt01}
implies that the formation of the stars in a present-day galaxy spheroid was 
fundamentally 
connected to the growth of the black hole as an active galactic nucleus 
(AGN). In fact, an evolutionary connection between the star-forming and 
AGN-dominated phases of galaxies has long been posited 
\citep[e.g.][]{sanders88}.

It is now known that
a significant fraction of cosmic star formation took place in
submillimetre galaxies (SMGs): dusty, ultraluminous, and highly
obscured galaxies at high redshift \citep{smail97, hughes98}. 
Although some distant QSOs have been detected as 
submillimetre sources in pointed observations 
\citep[e.g. ][]{mcmahon99,page01a}, the overlap between the 
QSO and SMG populations is relatively
small \citep{fabian00, severgnini00, almaini03, waskett03, lutz10}. 
In an attempt to uncover the relationship
between quasi-stellar objects (QSOs) and SMGs, \citet{page04} observed
matched samples of X-ray absorbed and unabsorbed QSOs at 850$\mu$m
with SCUBA.  These observations revealed a remarkable dichotomy in the
submillimetre properties of these two groups of sources: a substantial fraction
($\sim$ 50 per cent) of X-ray
absorbed QSOs are submillimetre galaxies, while X-ray
unabsorbed (i.e. normal) QSOs are not. This suggests that the two
types are linked by an evolutionary sequence, in which the X-ray
absorbed QSOs correspond to the transition between the main
star-forming phase and the QSO phase of a massive galaxy
\citep{page04,stevens05,coppin08}. The space density and luminosities of the
X-ray absorbed QSOs indicate that this transitional phase is
relatively brief, $\sim$15 per cent of the duration of the luminous
QSO phase.

However, the physical driver for this transition between obscured star-forming
galaxy and luminous QSO is unknown. Furthermore, the nature of the X-ray
absorption in the X-ray absorbed QSOs 
remains puzzling. These objects are characterised by hard, absorbed
X-ray spectra, but they have optical/UV spectra which are typical for
QSOs, with broad emission lines and blue continua. Assuming that their
hard X-ray spectral shapes result from photoelectric absorption from
cold material with solar abundances, the column densities are of order
$10^{22}$~cm$^{-2}$. The X-ray absorption could
be due to gas located within the AGN structure, or from more distant
material in the host galaxy.  Wherever the absorber is located, it
appears to contain very little dust: for a Galactic
gas/dust ratio, the restframe ultraviolet spectra would be heavily
attenuated by such large columns of material. 

Therefore in order to
investigate the X-ray absorption in these objects, 
we have obtained deep (50--100ks) \xmm\ observations of five X-ray
absorbed QSOs from our sample of hard-spectrum {\em Rosat} sources
\citep{page01b} that were detected as powerful submillimetre sources 
with SCUBA \citep{page01a, stevens05}. 

Throughout this paper we assume $H_{0}=70$~km~s$^{-1}$~Mpc$^{-1}$,
$\Omega_{m}=0.3$, and $\Omega_{\Lambda}=0.7$. We define a power law
spectrum such that $f_{\nu} \propto \nu^{-\alpha}$; we assign the
energy index $\alpha$ the subscript $X$ when it refers to the X-ray
spectral slope, $O$ when it refers to the optical/ultraviolet spectral slope and $OX$ when it refers to the optical to X-ray
spectral slope.

\section{Observations and data reduction}

\subsection{Optical spectra}

\begin{figure}
\begin{center}
\leavevmode
\psfig{figure=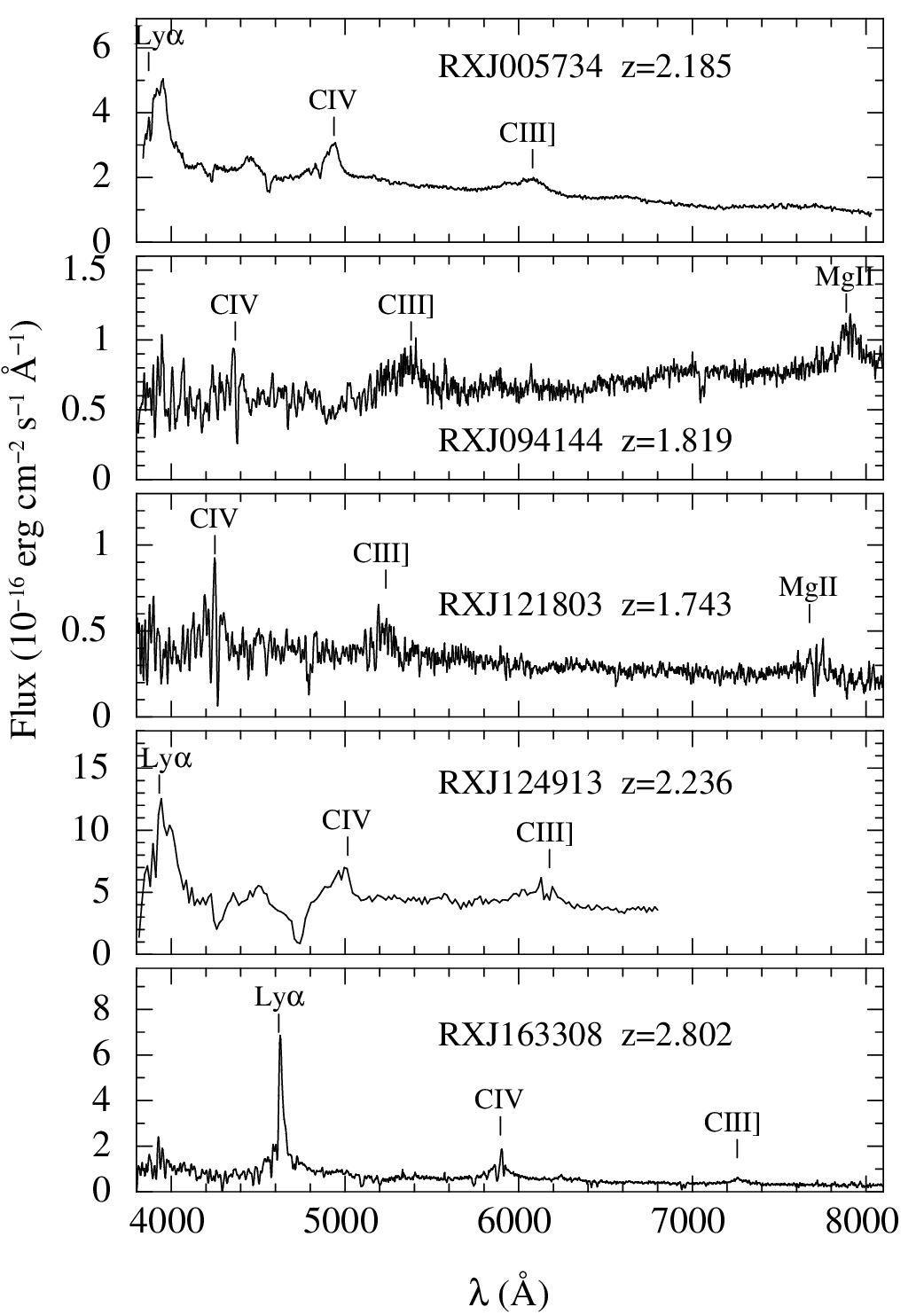,width=80truemm}
\caption{The optical spectra of \rxja, \rxjb, \rxjc, \rxjd\ and \rxje.}
\label{fig:optspecs}
\end{center}
\end{figure}

\rxja\ was observed on the 22nd November 1998, with the ESO 3.6m telescope 
at La Silla, Chile. The EFOSC~2 spectrograph was used with the 5000\AA\ blaze, 
300 lines mm\(^{-1}\) grism covering the wavelength range 3800\AA\ to
8000\AA, with a spectral resolution of 20\AA\ FWHM. 

\rxjb, \rxjc\ and \rxje\ were observed on the 3rd-5th March 1998 with the
William Herschel Telescope at the Observatorio del Roque de los Muchachos, La
Palma. The dual arm ISIS spectrograph was used with the 5400 \AA\ dichroic, the
R158R grating on the red arm and 
the R300B grating on the blue arm. The two ISIS arms combined give continuous
coverage of the wavelength range 3200\AA\ to 8100 \AA\ with resolution of 5\AA\
FWHM in the blue and 9\AA\ FWHM in the red.

In both observing runs, arc-lamp spectra and observations of 
spectrophotometric standard stars were taken for wavelength and 
flux calibration. The spectra were reduced and calibrated using 
standard routines in 
{\sc IRAF}.

\rxjd\ was not observed because it is a well known 
broad absorption line QSO (BALQSO) with optical spectra available in the 
literature; the spectrum shown in Fig. \ref{fig:optspecs} is taken 
from \citet{junkkarinen87}. 

\begin{table*}
\begin{center}
\caption{\xmm\ observations. Exposure times refer to the clean exposure times
  after periods of high background have been excluded. The UVW1 magnitudes 
for \rxjb\ and \rxjc\ come from the XMM-OM observations performed 
simultaneously with the X-ray observations, and are in the XMM-OM Vega 
system.}
\label{tab:observations}
\begin{tabular}{cccccccccc}
Target&Obs ID&date&\multicolumn{3}{c}{Filters}&\multicolumn{3}{c}
{Exposure time (ks)}&UVW1\\
      &   &    & pn & MOS1 & MOS2 & pn & MOS1 & MOS2&mag\\     
\hline\\
RX J005734.78$-$272827.4&0302310301&18 Dec 2005&thin&thin&thin&40.2&47.7&47.3&-\\
&&&&&&&&&\\
RX J094144.51$+$385434.8&0203270101&17 May 2004&medium&medium&medium&32.6&40.5&40.8& $21.63\pm 0.12$ \\
&&&&&&&&&\\
RX J121803.82$+$470854.6&0203270201&1 June 2004&thin&thin&thin&30.0&36.9&37.2& $20.63\pm 0.06$ \\
&&&&&&&&&\\
RX J124913.86$-$055906.2&0060370201&11 July 2001&thin&thin&thin&25.0&39.6&39.6&-\\
&0203270301&15 July 2004&medium&medium&medium&34.3&42.5&42.9&-\\
&&&&&&&&&\\
RX J163308.57$+$570258.7&0302310101&7 Sept 2005&thin&thin&thin&5.5&9.2&8.8&-\\
&0302310501&23 Oct 2005&thin&thin&thin&10.5&19.7&20.0&-\\
\end{tabular}
\end{center}
\end{table*} 

\subsection{X-ray spectra and ultraviolet images}

The \xmm\ observations used in this analysis are listed in Table
\ref{tab:observations}. The data were reduced with \xmm\ SAS version
6.5, and the calibration release of June 2006. After initial
processing of the EPIC MOS and pn data, spectra were extracted from a
circular region of 15 arcsecond radius, centred on the
target. Background was obtained from a circular region of radius 150
arcseconds, within which all significant sources were masked
out. Appropriate response matrices and effective area files were
generated with the {\sc SAS} tasks {\sc rmfgen} and {\sc arfgen}. For
each object, spectra and response matrices were combined using the
method described in \citet{page03}. To determine how the fluxes of the
sources compared to those obtained in our original {\em Rosat} survey,
we fitted the spectra below 2 keV with a powerlaw and fixed Galactic
absorption. For \rxja, \rxjc\ and \rxje\ we obtain $0.5-2$~keV fluxes of
$1.2\times 10^{-14}$, $1.5 \times 10^{-14}$ and $1.8\times
10^{-14}$~erg~cm$^{-2}$~s$^{-1}$ respectively, consistent with those
obtained with {\em Rosat} \citep{page00}. For \rxjb\ we 
obtain a $0.5-2$~keV flux
of $5\times 10^{-15}$~erg~cm$^{-2}$~s$^{-1}$, approximately a quarter
of its flux during our {\em Rosat} survey, while for \rxjd\ we obtain a
$0.5-2$~keV flux of $1.3\times 10^{-14}$~erg~cm$^{-2}$~s$^{-1}$, approximately
half the {\em Rosat} flux reported by \citet{page00}.

Simultaneous with the X-ray observations, the {\em XMM-Newton} Optical
Monitor (XMM-OM) took deep images in the UVW1 filter\footnote{UVW1
  images were not obtained during the July 2001 observation of \rxjd,
  in which the UV grism of XMM-OM was used instead.}, which has an
effective wavelength of 2910\AA. These observations were obtained
primarily to provide photometric redshift constraints for objects
surrounding the QSOs, and for the three highest redshift objects the
flux in this band is severely affected by the Lyman break. For \rxjb\
and \rxjc\ however, the Lyman break cuts in only at the extreme blue of the
UVW1 transmission, and will have a minor impact on the photometry. The
data were reduced using the XMM-OM SAS version 6.5. The UVW1 magnitudes of 
\rxjb\ and \rxjc\
(in the XM-OM Vega system) are given in Table~\ref{tab:observations}.

\section{Results}
\label{sec:results}

\subsection{Absorption in the rest-frame ultraviolet}
\label{sec:uvlines}

\begin{figure}
\begin{center}
\leavevmode
\psfig{figure=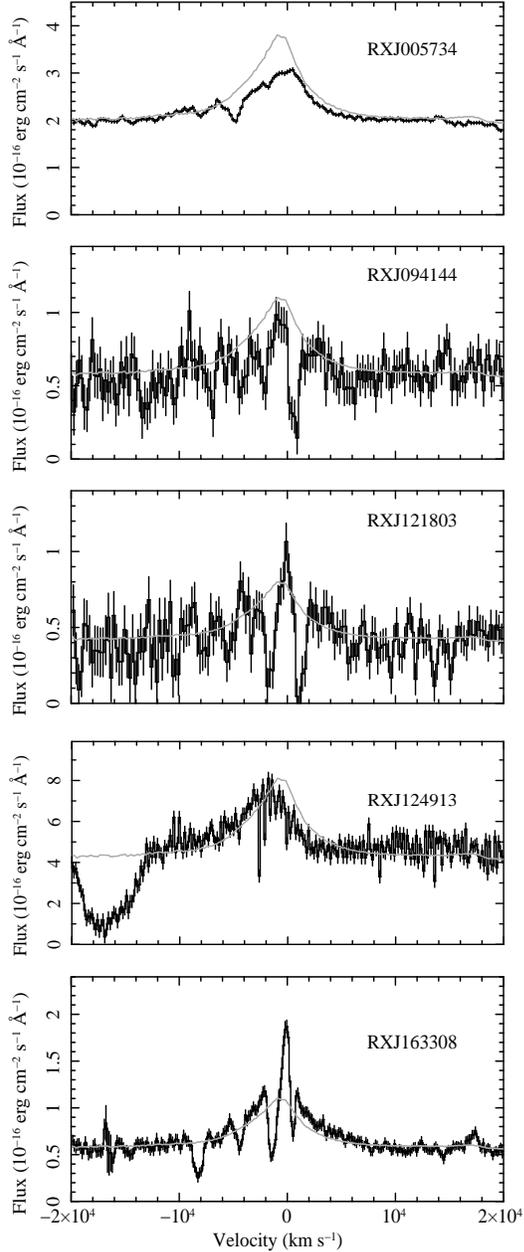,width=70truemm}
\caption{Close up of the region around the CIV 1550\AA\ emission line 
in our five X-ray absorbed QSOs. For comparison, the grey line shows the
median Sloan Digital Sky Survey QSO template from \citet{vandenberk01}}.
\label{fig:CIV}
\end{center}
\end{figure}

Broadband optical spectra of the five X-ray absorbed QSOs are shown in
Fig. \ref{fig:optspecs}. It is notable that in all five objects, at least one
absorption line is superimposed on the broad C\,IV emission line 
or on the continuum to the blue of this line. To examine
this more closely, we show the regions around the C\,IV emission line in more
detail in Fig. \ref{fig:CIV}. In \rxja, there are two absorption lines
apparent, with outflow velocities of 4800 and 8100 km~s$^{-1}$, assuming that
they are C\,IV. In the spectrum of
\rxjb, at least one absorption line is present, apparently inflowing with a
velocity of 700~km~s$^{-1}$ assuming that it is C\,IV. As this spectrum is
relatively noisy, we are unable to confirm whether the tentative absorption 
features to the blue of C\,IV are real or statistical fluctuations. 
There are two significant absorption features in \rxjc, one inflowing at 
1200~km~s$^{-1}$, and one outflowing at 1600~km~s$^{-1}$. Mg\,II absorption is
detected from the inflowing component.
For \rxjd\ we
adopt an emission line redshift of $z=2.236$ \citep{hewitt93}, although
the reported emission line redshift of this source ranges from $z=2.22$
\citep{boksenberg78} to
$z=2.244$ \citep{hill93} 
depending on which emission lines the redshift is based on. Relative to this
emission line redshift, the C\,IV broad absorption line has a central velocity
of -16700~km~s$^{-1}$, with a FWHM of 5700~km~s$^{-1}$ \citep{boksenberg78}.
In \rxje\ there are four absorption lines visible, but two of these
are likely to be due to an intervening absorption line system with $z=1.482$,
for which a number of other transitions are detected in the optical
spectrum. The remaining two lines have outflow velocities of 800 and 7700
km~s$^{-1}$. Both are accompanied by Ly$\alpha$ absorption with consistent
velocities, and absorption from  Si\,IV is detected in 
the 7700
km~s$^{-1}$ component. The C\,IV absorption line properties of the five QSOs 
are listed in Table~\ref{tab:uvlines}.

The five QSOs have strong rest-frame ultraviolet continuum
emission. Ultraviolet spectral slopes were measured from the optical
spectra, longward of Ly$\alpha$, after masking emission and absorption
lines. The ultraviolet spectral slopes are listed in
Table~\ref{tab:uvlines}.  With the exception of \rxjb, the ultraviolet
continua of the five QSOs have spectral slopes in the range
$0<\alpha_{O}<1$, which is typical for optically-selected QSOs
\citep{francis91}, indicating little or no dust reddening intrinsic to
these objects. On the other hand, as can be seen in
Fig. \ref{fig:optspecs} the rest-frame ultraviolet continuum of \rxjb\
is much redder than those of the other objects. The continuum slope in
this object is $\alpha_{o}=2.7$, outlying from the distribution of
optically-selected QSOs at the 3$\sigma$ level \citep{francis91}, and
suggesting that this object may be somewhat dust reddened. The
ultraviolet spectral shape can be matched by reddening the median
Sloan Digital Sky Survey QSO spectrum \citep{vandenberk01} by
E(B$-$V)=0.20 using the SMC reddening law \citep{pei92}, 
a similar level of extinction as observed in the nearby 
starburst/BALQSO IRAS~$07598+6508$ 
\citep{lipari94}, and suggesting 
$\sim$1.4 magnitudes of extinction at 2500\AA\ in the restframe. The UVW1$-$B
colour of \rxjb\ (where the B magnitude comes from \citealt{page01b})
is in reasonable agreement with this reddened template: after
correction for Galactic extinction the UVW1-B colour of \rxjb\ is 
$0.47\pm 0.18$
compared to UVW1$-$B=0.71 predicted by the reddened template. 
In contrast, \rxjc\ has a
much bluer UVW1$-$B=$-0.45\pm 0.11$ after correction for Galactic extinction. 
For comparison, the median SDSS QSO template would have UVW1$-$B=$-0.68$ at 
the redshift of \rxjc.

\begin{table}
\begin{center}
\caption{Characteristics of 
C\,IV ultraviolet absorption line systems and ultraviolet
  continuum slopes $\alpha_{o}$. Negative velocities
  correspond to blueshifted lines and positive velocities correspond to
  redshifted lines. 
}
\label{tab:uvlines}
\begin{tabular}{ccccc}

QSO&Velocity     &C\,IV~FWHM&C\,IV~EW&$\alpha_{o}$\\
   &(km~s$^{-1}$)&(km~s$^{-1}$)     &(\AA)       &\\
\hline\\
\rxja& -4800  & 500  & 5  & 0.6 \\
\rxja& -8100  & 500  & 2  & 0.6 \\
&&&&\\
\rxjb&  700   & 800  & 12 & 2.7 \\
&&&&\\
\rxjc&  1200  & 900  & 15 & 1.0 \\
\rxjc& -1600  & 900  & 14 & 1.0 \\
&&&&\\
\rxjd& -16700 & 5700 & 70 & 1.0 \\
&&&&\\
\rxje& -800   & 700  & 11 & 0.1 \\
\rxje& -7700  & 700  & 10 & 0.1 \\
&&&&\\
\end{tabular}
\end{center}
\end{table}

\begin{figure}
\begin{center}
\leavevmode
\psfig{figure=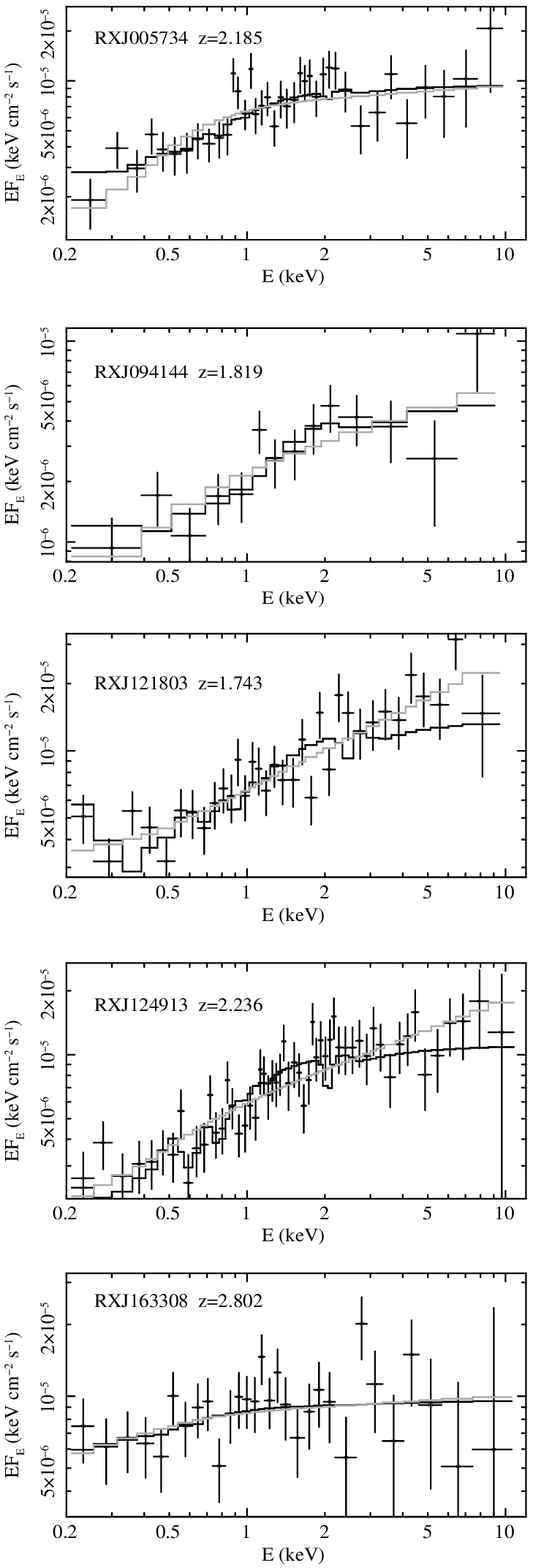,width=70truemm}
\caption{EPIC spectra of the X-ray absorbed QSOs together with the 
$\alpha_{X}=0.98$
power law and ionized absorber model (black stepped line) and free-$\alpha$ 
power law and
cold absorber model (grey stepped line). All spectra are shown in the observer
frame. Both model and data have been divided by the product of the effective
area and Galactic transmission as a function of energy.}
\label{fig:epicspectra}
\end{center}
\end{figure}

\subsection{Modelling the X-ray spectra}
\label{sec:modelling}

Analysis of the X-ray spectra was performed using SPEX version 
2.00.11\footnote{http://www.sron.nl/divisions/hea/spex/index.html}.  As a
starting point, the \xmm\ spectra were fitted with a similar
model to that considered by \citet{page01b}: a power law with a fixed
$\alpha_{X}$ of 0.98, fixed Galactic absorption, and cold photoelectric
absorption at the QSO redshift with $\nh$ as a free parameter. The
value of $\alpha_{X}=0.98$ was chosen because it is the mean spectral index of 
QSOs in the
\xmm\ energy range and at similar X-ray flux levels to our
targets, as determined by \citet{mateos05a} and confirmed 
in more recent work \citep{mateos10}. The results are given in 
Table \ref{tab:modelfits}. Although some of the individual objects are 
successfully fit by this model, others are not (\rxjc\ and \rxjd), and 
 the total $\chi^{2}/\nu=225/161$ for the sample of 5 objects is not acceptable.
Therefore, we attempted to fit the spectra with two models, one in which 
the continuum spectral slope was allowed to vary, and one in which the 
ionization parameter of the absorber was allowed to vary.

\begin{table*}
\begin{center}
\caption{Model fits to the \xmm\ spectra. The models are a combination of a
  power law (PL), and either a cold photoelectric absorber (CA) or warm, 
ionized 
absorber (WA) at the redshift of the AGN. In addition, a zero-redshift cold 
absorber with column density fixed at the
  Galactic value is included in each model. Parameters marked 
`{\scriptsize f}' are fixed in the fit, and `*' is used to indicate an 
uncertainty in a model parameter that is not constrained by the fit. $\nh$ has 
units of $cm^{-2}$ and $\xi$ has units of erg\,cm\,s$^{-1}$.}
\label{tab:modelfits}
\begin{tabular}{lcccccc}
&\rxja&\rxjb&\rxjc&\rxjd&\rxje&Total\\
\hline\\
PL$\times$CA&&&&&&\\
$\alpha_{X}$&0.98$^{f}$&0.98$^{f}$&0.98$^{f}$&0.98$^{f}$&0.98$^{f}$&\\
$\log \nh$ &22.0$^{+0.1}_{-0.1}$&22.1$^{+0.2}_{-0.2}$&21.8$^{+0.1}_{-0.1}$&22.2$^{+0.1}_{-0.1}$&21.5$^{+0.2}_{-0.4}$&\\
$\alpha_{OX}$&1.70$^{+0.05}_{-0.06}$&1.48$^{+0.30}_{-0.12}$&1.32$^{+0.06}_{-0.05}$&1.89$^{+0.09}_{-0.09}$&1.44$^{+0.05}_{-0.04}$&\\
$\chi^{2}/\nu$&41/36&15/12&64/32&82/53&22/28&225/161\\
&&&&&&\\
PL$\times$CA&&&&&&\\
$\alpha_{X}$&0.82$^{+0.10}_{-0.09}$&0.57$^{+0.19}_{-0.17}$&0.41$^{+0.07}_{-0.06}$&0.53$^{+0.07}_{-0.07}$&0.94$^{+0.15}_{-0.13}$&\\
$\log \nh$ &21.8$^{+0.2}_{-0.2}$&21.5$^{+0.4}_{-0.9}$&0.00$^{+21.9}_{-*}$&21.5$^{+0.2}_{-0.3}$&21.4$^{+0.4}_{-*}$&\\
$\alpha_{OX}$&1.71$^{+0.05}_{-0.06}$&1.54$^{+0.52}_{-0.13}$&1.39$^{+0.07}_{-0.06}$&1.96$^{+0.10}_{-0.10}$&1.43$^{+0.05}_{-0.04}$&\\
$\chi^{2}/\nu$&39/35&11/11&29/31&50/52&22/27&151/156\\
&&&&&&\\
PL$\times$WA&&&&&&\\
$\alpha_{X}$&0.98$^{f}$&0.98$^{f}$&0.98$^{f}$&0.98$^{f}$&0.98$^{f}$&\\
$\log \xi$&2.4$^{+0.2}_{-0.3}$&2.7$^{+0.4}_{-0.4}$&2.9$^{+0.2}_{-0.3}$&2.9$^{+0.1}_{-0.1}$&2.3$^{+1.0}_{-6.3*}$&\\
$\log \nh$ &22.7$^{+0.2}_{-0.2}$&23.2$^{+0.4}_{-0.4}$&23.1$^{+0.3}_{-0.3}$&23.1$^{+0.1}_{-0.1}$&22.2$^{+1.0}_{-0.7}$&\\
$\alpha_{OX}$&1.68$^{+0.05}_{-0.06}$&1.43$^{+0.17}_{-0.10}$&1.26$^{+0.04}_{-0.04}$&1.85$^{+0.08}_{-0.08}$&1.43$^{+0.05}_{-0.04}$&\\
$\chi^{2}/\nu$&34/35&9/11&43/31&51/52&21/27&158/156\\
&&&&&&\\

\end{tabular}
\end{center}
\end{table*}

As seen in Table \ref{tab:modelfits}, allowing the continuum spectral
indices or the ionization parameters of the absorbers to vary produced
acceptable values of $\chi^{2}/\nu$ for the 5 objects individually,
and as a sample. Since the two models are equivalent in terms of
$\chi^{2}/\nu$, we can only discriminate between the two models by
looking to see which provides the more physically plausible and
self-consistent explanation for the spectra. In particular, we can
test the hypothesis that the underlying spectra of these objects are
reasonably typical for QSOs, and their unusual observed spectral
shapes are due to absorption. Two properties in particular are
important in this context: the spectral index of the underlying X-ray
spectrum, and the ratio of optical to X-ray flux, which we
parameterise as the equivalent $\alpha_{OX}$, which is the energy index of the
power law that would connect the restframe 2500\,\AA\ and 2 keV flux
densities.

Beginning with the X-ray spectral slope, the most appropriate
comparison sample of normal QSOs is the sample examined by
\citet{mateos05a}, which is an X-ray selected sample with a similar
flux limit to the \ros\ hard spectrum sample from which our targets
were drawn, and measured in the same energy range, and with the same
instrumentation (\xmm\ EPIC) as our targets. \citet{mateos05a} found
that the distribution of QSO spectral slopes can be described as a
Gaussian with a mean of $\alpha_{X}=0.98$ and a standard deviation
$\sigma_{\alpha}=0.21$. From Table \ref{tab:modelfits} we see that when
the spectral slopes are allowed to vary, the best fit values of
$\alpha_{X}$ for \rxja\ and \rxje\ are relatively close to the mean of
the $\alpha_{X}$ distribution. However, the best fit values of
$\alpha_{X}$ for \rxjb, \rxjc, and \rxjd\ are all unusually low, lying
within the lowest 3 percent of the distribution.  This is an unlikely
circumstance if the intrinsic photon indices of our sample were drawn
from the same distribution as QSOs in general, and suggests that the
cold photoelectric absorption does not fully account for the unusual
X-ray spectral shapes observed in these objects. In comparison, the
spectral slopes are fixed in the ionized absorber model, and so are by
design consistent with a typical underlying QSO spectrum.

Moving to $\alpha_{OX}$, we have calculated this quantity for each of
the model fits and each of the QSOs. In each case the restframe
2500\,\AA\ flux has been estimated from the B magnitude listed in
\citet{page01b}, assuming an optical slope of $\alpha_{O}=0.5$. The
2~keV flux is determined from the continuum model of the X-ray
spectral fit, i.e. it is the intrinsic continuum flux, corrected for
the modelled X-ray absorption. 

\begin{figure}
\begin{center}
\leavevmode
\psfig{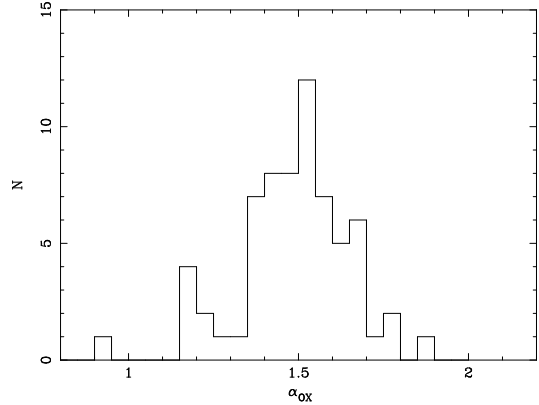}
\caption{Reference $\alpha_{OX}$ distribution for $z>1$ RIXOS 
X-ray selected QSOs, as described in Section \ref{sec:modelling}.}
\label{fig:rixosalphaox}
\end{center}
\end{figure}

The distribution of $\alpha_{OX}$ in normal QSOs is known to depend on
luminosity \citep{vignali03,strateva05} and there is inevitably a
selection bias depending on whether the sample is X-ray or optically
selected. The most appropriate reference distribution of $\alpha_{OX}$
in the literature is that presented in \citet{strateva05} for a sample
of 155 Sloan Digital Sky Survey QSOs with medium-deep \ros\
coverage. However, since this distribution is based on an
optically-selected sample, while our sample is X-ray selected, we have
constructed a reference distribution of $\alpha_{OX}$ for X-ray
selected QSOs drawn from the \ros\ International X-ray Optical Survey
\citep[RIXOS; ][]{mason00}. We restrict the reference sample to those
RIXOS QSOs at $z>1$, which corresponds to a similar luminosity range
to our X-ray-absorbed QSO sample, because RIXOS has a similar X-ray
flux limit to the \ros\ hard spectrum survey from which our targets
were drawn. We calculated $\alpha_{OX}$ for the reference sample in an
identical fashion to our absorbed QSO sample except that we assume
no intrinsic X-ray absorption and $\alpha_{X}=0.98$ for the RIXOS
AGN. The reference sample has a mean $\langle \alpha_{OX}\rangle =
1.49$ and a standard deviation $\sigma_{OX} = 0.16$; the distribution
of $\alpha_{OX}$ for the reference sample is shown in
Fig. \ref{fig:rixosalphaox}. For comparison, the SDSS sample presented
in \citet{strateva05} has a very similar distribution of $\alpha_{OX}$,
with $\langle \alpha_{OX}\rangle = 1.48$ and $\sigma_{OX} = 0.18$.

In Table \ref{tab:modelfits} we see that with the exception of \rxjd,
the absorbed QSOs have $\alpha_{OX}$ values within two standard
deviations of $\langle \alpha_{OX}\rangle$ for the reference
distribution. For \rxjd, the level of discrepancy depends strongly on
the X-ray spectral model. The ionized absorber fit gives the most
consistent $\alpha_{OX}$ value, at 2.3 standard deviations from the
mean of the reference sample, while the fit with a cold absorber and
fitted $\alpha_{X}$ gives the most discrepant value of $\alpha_{OX} =
1.96$, $> 2.9$ standard deviations from the mean of the reference
sample. Indeed, this value of $\alpha_{OX}$ is larger than any
$\alpha_{OX}$ in the RIXOS reference sample or the SDSS sample of
\citet{strateva05}, suggesting that the model with a cold absorber and
the continuum slope fitted as a free parameter would also require a
highly unusual spectral energy distribution for \rxjd. While \rxjb\
has rather typical values of $\alpha_{OX}$ in
Table~\ref{tab:modelfits}, we note that 
if we correct for the intrinsic extinction
equivalent to E(B$-$V)=0.2, suggested by the UV continuum shape
(Section~\ref{sec:uvlines}), the $\alpha_{OX}$ would increase by 0.40,
leading to an abnormal $\alpha_{OX}$ for the cold absorber, fitted
$\alpha_{X}$ model for this source as well as for \rxjd.

Overall, the ionized absorber model is consistent with the five QSOs
having unremarkable underlying X-ray continua and optical to X-ray
spectral energy distributions. In contrast, if the X-ray absorbers are
modeled as cold gas, the QSOs require a combination of cold
absorption, unusual X-ray continua, and unusual optical to X-ray
spectral energy distributions. We therefore consider that the ionized
absorber model provides a simpler, less contrived solution than the
model in which the X-ray absorber is cold. We also note that the
presence of C\,IV absorption lines with EW$>5$\,\AA\ 
in the restframe
UV spectra of all five QSOs (and Si\,IV in \rxjd\ and \rxje) 
provides independent confirmation that
these objects are viewed through columns of ionized gas.  Therefore in
what follows we adopt the ionized absorber fits as the best
description of the X-ray spectra.

\begin{figure*}
\begin{center}
\leavevmode
\psfig{figure=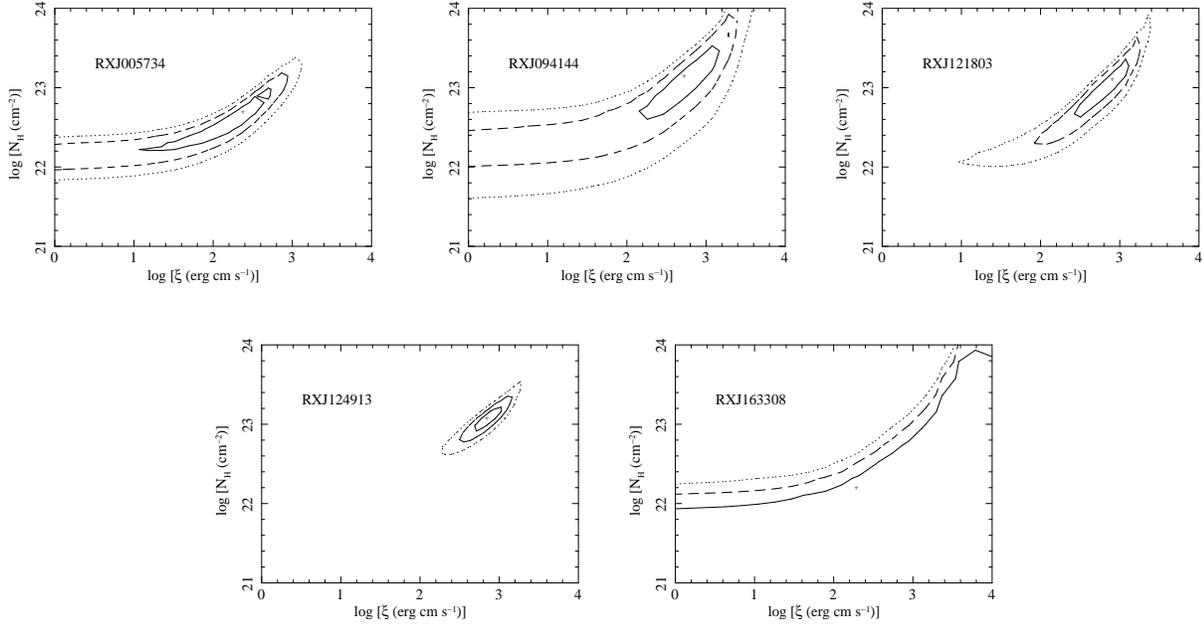,width=160truemm}
\caption{Confidence contours for $\xi$ and $\nh$ in the ionized absorber 
model fits to the \xmm\ spectra.}
\label{fig:xinhcontours}
\end{center}
\end{figure*}

\section{Discussion}
\label{sec:discussion}

\subsection{Characteristics of the absorbers}
\label{sec:characteristics}

Our study has revealed absorption from ionized gas in both the
rest-frame UV spectra and in the X-ray spectra of our sample of X-ray
absorbed QSOs. In the nearby Universe, ionized absorbers are common in
Seyfert galaxies \citep[e.g. ][]{reynolds97,george98}, and have been
studied in some detail \citep{blustin07,steenbrugge05}.  Therefore in
trying to understand the properties of our sample of X-ray absorbed
QSOs, it is instructive to compare them with those found in
Seyferts. Amongst Seyfert galaxies, ionized absorption is observed
primarily in broad-line, (type 1) objects 
\citep[e.g.][]{reynolds97,george98,blustin05};
indicating that the ionized absorbers lie within the ionization cones
of the AGN, assuming the standard AGN geometric
unification scheme in which the central regions are surrounded by a
dusty, obscuring torus \citep{antonucci93}.  
The presence of ionized absorption within our X-ray
absorbed QSOs is therefore in keeping with a normal pole-on
orientation for these objects.

The ionized column densities, and the degree of soft X-ray
attenuation, of our X-ray absorbed QSOs are considerably higher than
are found in the majority of nearby Seyferts: with the exception of
\rxje\ for which the X-ray absorber properties are poorly constrained,
all of our sample have best fit log~$\nh$ $>22.5$, whereas only 3 of
the 23 AGN in the sample studied by \citet{blustin05}, have log~$\nh$
$>22.5$. Nonetheless, a few nearby objects such as NGC3783 do have
comparable column densities and ionization states
\citep{blustin02,netzer03,reeves04,krongold03}.  The ionized absorbers
in Seyferts typically contain ions covering a large range of
ionization, and where good data are available they almost always
require multiple phases with different ionization parameters to model
their spectra \citep{blustin05,mckernan07}.  Moving to higher
luminosity objects, ionized absorbers appear to be common in
QSOs, with the studies of \citet{porquet04} and \citet{piconcelli05}
suggesting that around 50 per cent of low redshift QSOs show some
evidence for ionized absorption in their X-ray spectra. However, 
column densities of ionized material similar
to those found in our sample of X-ray absorbed QSOs are relatively
rare, with \citet{piconcelli05} finding that only 2 of
their sample of 40 nearby QSOs have ionized absorbers with best fit
log~$\nh$ $>22.5$.  In the small number of nearby QSOs which have
well-characterised ionized absorbers with $\nh>
10^{22}$\,cm$^{-2}$, notably PG\,1114+445 \citep{ashton04},
PDS\,456 \citep{reeves03}, H\,$0557-385$ \citep{ashton06}, and
PG\,1211+143 \citep{pounds03}, the absorbers are found to have
multiple ionization phases, in common with the well-studied Seyfert warm
absorbers.

On this basis, we can expect that the parameters determined for the single 
phase absorbers in
our sources are therefore likely to be representative only of the dominant
phase of absorption, and the high-$\xi$ absorption which we have detected is
likely to be accompanied
by significant soft X-ray absorption due to material with log $\xi <
1.5$. In fact the multi-ionization-phase 
nature of the ionized absorbers is implied by the
absorption lines detected in their UV spectra. The best fit ionized X-ray 
absorber
models for all 5 objects produce insignificant C\,IV absorption lines (EW $<
1$\,\AA), in contrast to the large EW lines ($> 5$\,\AA) that are 
measured in their rest-frame UV spectra, implying that at least two 
ionization phases of absorber are required to produce both UV and X-ray 
absorption. 
Note that while absorption from C\,IV is seen in about half of 
QSOs \citep{vestergaard03}, C\,IV absorption lines with equivalent
widths $> 5$\,\AA\ are relatively rare. Only 3 out of 114 non-BAL
QSOs studied by \citet{vestergaard03} show C\,IV absorption with
such large EW. Combined with the statistic that $\sim$ 15 percent of
QSOs have C\,IV BALs, which by definition have C\,IV EW $> 5$\,\AA\
\citep{gibson09,tolea02} we can conservatively estimate the
incidence of C\,IV absorption lines with EW$> 5$\,\AA\ to be 20 per
cent. The probability of all 5 of our X-ray absorbed QSOs having
such absorption lines if they were drawn randomly from the
population is therefore $0.2^{5}=3\times 10^{-4}$, and hence the
connection of the C\,IV absorbers with the X-ray absorption is quite
secure.

Our finding that the X-ray absorbers in our sample are highly ionized
allows us to address the puzzling absence of strong UV attenuation by
dust, despite the large column densities of X-ray absorbing
gas. Nearby Seyfert 1 galaxies with warm absorbers also show this
property: the column densities of X-ray absorbing gas can be large but
they have bright ultraviolet continua with broad emission lines,
e.g. NGC\,3783 has $\log \nh > 22.5$ \citep{netzer03,reeves04} as well
as a strong UV continuum and broad UV emission lines with little
reddening \citep{crenshaw01,evans04}.  

As a corollary to the apparent similarity between the ionized
  absorbers found in our X-ray absorbed QSOs and some nearby Seyferts,
  we might wonder whether star formation is associated with such X-ray
  absorption in objects such as NGC\,3783, as it is in X-ray absorbed
  QSOs. A systematic investigation of this issue is beyond the scope
  of this paper, but it is interesting to note that NGC\,3783 has a
  2-10 keV X-ray luminosity of $1.3\times 10^{43}$~ergs~s$^{-1}$ and
  an infrared luminosity of $3 \times 10^{10}$~L$_{\odot}$
  \citep{mullaney11}, so that if we scaled its SED to match the X-ray
  luminosities of our X-ray absorbed QSOs its infrared luminosity
  would be in the regime of ultraluminous infrared galaxies,
  commensurate with a substantial star formation rate.

\subsection{Comparison to soft X-ray weak QSOs}
\label{sec:xrayweak}

It is interesting to compare the X-ray absorbed QSOs with another
population of unusual QSOs: those that are classified as soft X-ray
weak. These objects are characterised by large values of
$\alpha_{ox}$, and represent around 10\% of optically selected QSOs
\citep{laor97}. \citet{brandt00} found a strong correlation between
$\alpha_{ox}$ and the equivalent width of C\,IV absorption, which is
found to be $> 4$\AA\ in the majority of soft X-ray weak QSOs. This
was interpreted by \citet{brandt00} as evidence that soft X-ray weak
QSOs are such because they are absorbed in the soft X-ray band.  As
shown in Section \ref{sec:uvlines}, the QSOs in our sample all have
C\,IV absorption lines with EW$>4$\AA, similar to the C\,IV lines
found in soft X-ray weak QSOs. More recently, \xmm\ observations of
soft X-ray weak QSOs have indicated that this absorption is primarily
due to ionized gas \citep{schartel05,brinkmann04,piconcelli04}, with
ionization parameters ranging from log~$\xi=1.9$ to log~$\xi=3.2$ and
column densities ranging from $3 \times 10^{22}$~cm$^{-2}$ to $2\times
10^{23}$~cm$^{-2}$. These column densities and ionization parameters
are remarkably similar to those of our X-ray absorbed QSOs. 
 Given these similarities, we might expect that intense star
  formation would be as common in soft X-ray weak QSOs as it is in X-ray
  absorbed QSOs. To our knowledge there has been no systematic study
  of the star formation rates in soft X-ray weak QSOs, but there are
  certainly well known examples of soft X-ray weak QSOs which are
  rapidly forming stars, notably Mrk~231 and IRAS~$07598+6508$
  \citep{lipari94}.

\subsection{Location of the absorbers}
\label{sec:location}

We now consider the location, physical nature, and implications of the ionized
absorbers found in our sample. To estimate the location we
assume that the
absorber is in the form of an outflowing wind originating at a distance $R_{abs}$ 
from
the continuum source with density $n \propto r^{-2}$. From the definition of
the ionization parameter 
\begin{equation}
\xi =
L_{ion}/(nr^{2})
\label{eq:xi}
\end{equation} 
where $L_{ion}$ is the 1-1000 Rydberg ionizing 
luminosity in
erg~s$^{-1}$, such a wind will have an ionization parameter 
$\xi$ that is
constant with radius $r$. To take account that the winds from AGN are
likely to be clumpy or
filamentary, we characterise the outflow with a macroscopic volume filling
factor $f<1$. The column
density of the absorber is dominated by material close to the base of the
outflow, and hence $R_{abs}$ can be taken as the characteristic distance of the
absorber from the continuum source. Integrating the density along the line of
sight, we obtain $\nh = fR_{abs}n_{abs}$ where $n_{abs}$ is the density at the
base of the outflow. Substituting $n_{abs}$ for $n$ in Equation \ref{eq:xi}, 
we obtain
the relation 
\begin{equation}
R_{abs}=\frac{L_{ion} f}{\xi \nh}
\label{eq:rabs}
\end{equation}
To
estimate $L_{ion}$ we assume that below 0.35 keV the SED has the functional form
given by \citet{mathews87} for a QSO ionizing spectrum, in which the X-ray 
power law steepens to $\alpha=3$ in the range 56--350 eV, has a slope of
$\alpha=1.0$ in the 24--56 eV range, below which it has a 
slope of $\alpha_{O}=0.5$. This SED of \citet{mathews87} was constructed 
to be appropriate for objects of QSO luminosity, and when joined to our 
$\alpha_{X}$=0.98 spectrum at 0.35 keV it has $\alpha_{OX}=1.47$. 

For the fixed
$\alpha_{X}$=0.98 which we have taken in our ionized absorber model
fits, $L_{ion}$ is a factor of 44.3 times larger than the restframe
2-10 keV luminosity. For the filling factor $f$ we assume a value of
$\log f = -2$, which is the average $\log f$ for ionized absorbers
with $\xi>0.7$ in Seyfert galaxies found by \citet{ashton06}. The
derived values of $L_{ion}$ and $R_{abs}$ are given in Table
\ref{tab:physics}. With the exception of \rxje, for which $R_{abs}$ 
is very poorly constrained, 
the best fit values of $R_{abs}$ range from 1 to 15 parsecs. 
The uncertainties given for $R_{abs}$ are derived
from the statistical uncertainties on the product $\xi \times \nh$,
and so do not include the uncertainty on the volume filling factor. As
an estimate of this additional uncertainty, we note that \citet{ashton06} 
obtained a standard deviation in $\log f$ of 0.8 for Seyfert warm 
absorbers with $\xi>0.7$. For comparison, we also provide in Table
\ref{tab:physics} the locations of the inner edge of the dusty torus
$R_{torus}$ for our sources, calculated using equation 5 from
\citet{barvainis87}, with a dust sublimation temperature of 1260\,K 
\citep[as implied by the composite near-IR QSO spectrum obtained by ][]{glikman06} and taking the ultraviolet luminosity to be
equal to $L_{ion}$.
 We see that 
$R_{abs}$ is in all cases smaller than, or consistent with, $R_{torus}$.

It would therefore be natural to assume that the absorbers originate
in the AGN themselves, rather than in their host galaxies. This
solution is attractive, because it is compatible with the lack of
optical extinction in these objects: if the absorber is driven as a
wind, either from the accretion disc or from evaporation of the inner
edge of the molecular torus, then dust will be sublimated before (or
as) it enters the flow.  With this in mind, it is instructive to
obtain upper limits for the distance of the absorbing gas from the
nucleus at which it could remain ionized by the AGN radiation field,
using Equation \ref{eq:rabs} and taking the limit case of filling
factor $f=1$. Doing so, we find that the X-ray absorbers in \rxjb,
\rxjc\ and \rxjd\ must lie within a few hundred parsecs of the
ionizing source, within or close to the AGN rather than being
distributed throughout the host galaxy.

\begin{table}
\begin{center}
\caption{Characteristic scales and outflow rates for the sample of X-ray
absorbed QSOs. $L_{ion}$ is the 1-1000 Rydberg ionizing luminosity in
erg~s$^{-1}$, $R_{abs}$ is the distance of the base of the absorber from the
continuum source in cm, $R_{torus}$ is the distance of the inner edge of the
dusty torus from the continuum source in cm, {\it \.{M}}$_{acc}$ is the mass
accretion rate and {\it \.{M}}$_{out}$ is the mass outflow rate in the
ionized absorber. An `$*$' indicates a confidence limit that is not
constrained by the data. Note that the uncertainties are purely 
statistical, and do not include the systematic uncertainty on the 
filling factor $f$, on which both $R_{abs}$ and {\it \.{M}}$_{out}$ 
depend linearly. As explained in Section \ref{sec:location}, we 
estimate an uncertainty on log~$f$ of $\sim0.8$.}
\label{tab:physics}
\begin{tabular}{l@{\hspace{1mm}}c@{\hspace{2mm}}c@{\hspace{2mm}}c@{\hspace{2mm}}c}
Object&$log L_{ion}$&$\log R_{abs}$&$\log R_{torus}$&
$\log (${\it \.{M}}$_{out}/${\it \.{M}}$_{acc})$\\
&&&&\\
\rxja&46.7&$19.6^{+0.7}_{-0.4}$& 19.4&$1.1^{+0.4}_{-0.3}$\\
\rxjb&46.2&$18.4^{+0.6}_{-0.8}$& 19.2&---\\
\rxjc&46.6&$18.6^{+0.5}_{-0.6}$& 19.4&$-0.1^{+0.4}_{-0.2}$\\
\rxjd&46.8&$18.8^{+0.3}_{-0.2}$& 19.5&$0.9^{+0.2}_{-0.1}$\\
\rxje&47.0&$20.7^{+*}_{-2.4}  $& 19.5&$1.2^{+*}_{-1.2}$\\
\end{tabular}
\end{center}
\end{table}

A key indicator of the importance of an ionized AGN wind is the ratio
of the mass outflow rate {\it \.{M}}$_{out}$ to the accretion rate {\it
\.{M}}$_{acc}$. To determine this ratio, we first take $L_{ion}$ to be
$0.6\times L_{acc}$ \citep{vasudevan09} where $L_{acc}$ is 
the bolometric accretion 
luminosity, so that
\begin{equation}
L_{ion}=0.6L_{acc}=0.6\epsilon${\it \.{M}}$_{acc} c^{2} 
\label{eq:ion}
\end{equation}
where $c$ is the speed of
light, and $\epsilon$ is the accretion efficiency.
 We can obtain the mass outflow
rate of the wind using equation 18 from \citet{blustin05}:
\begin{equation}
${\it \.{M}}$_{out}=\frac{1.23 m_{p} L_{ion} f v \Omega}{\xi}
\label{eq:blustin}
\end{equation}
where $m_{p}$ is the rest mass of a proton and $v$ is the outflow velocity, and
 $\Omega$ is the solid angle of the outflow. Combining equations \ref{eq:ion}
 and \ref{eq:blustin} we obtain:
\begin{equation}
${\it \.{M}}$_{out}/${\it \.{M}}$_{acc} = 0.74 m_{p} f v 
\Omega \epsilon c^{2} / \xi
\label{eq:outflow}
\end{equation}
We
take $\epsilon=0.1$ and as before we assume $f=0.01$. We take the opening angle
of the outflow $\Omega = \pi$ steradians, corresponding to the
opening angle of the torus inferred from the type 1 / type 2
number ratio in nearby Seyferts \citep{maiolino95}.
 For the velocity $v$ we
assume that the X-ray absorbing ionized outflow is associated with the fastest
outflow identified in the restframe ultraviolet (Table \ref{tab:uvlines}). This
is a reasonable estimate, since the highest ionization absorbers are typically 
(though not always)
the highest velocity systems in individual Seyfert galaxies
\citep[e.g. ][]{sako03, steenbrugge05, constantini07}. At the same time it is a
conservative estimate (i.e. may underestimate the mass outflow rate) for two
reasons: firstly, the signal to noise ratios of our optical spectra
(particularly for \rxjb\ and \rxjc) are not high enough that we can be
confident we have identified all of the absorption line systems, and secondly
our estimates of the systemic velocities are based on the UV broad emission
lines, which themselves are frequently observed to have a net outflow velocity
of order 1000 km~s$^{-1}$ \citep{espey89}. The derived values of log~{\it
\.{M}}$_{out}/${\it \.{M}}$_{acc}$ are listed in Table \ref{tab:physics}.  In
the case of \rxjb, the C\,IV absorption line that is identified in the rest-frame UV
spectrum is redshifted with respect to the UV emission lines, so its true
velocity with respect to the system is too uncertain for us to attempt to
estimate the mass outflow rate. The mass outflow rates derived for 
\rxja, \rxjd\ and \rxje\ are about 10 times their mass accretion rates, while
in \rxjc\ the accretion and outflow rates are comparable.

\subsection{Implications for QSO evolution}
\label{sec:evolution}

At this point it is worth considering how these results bear on the
broader picture of how X-ray absorbed QSOs relate to the evolution of
massive black holes and their host galaxies. In \citet{page04} we
argued that the submillimetre properties of X-ray absorbed QSOs
imply an evolutionary sequence in which submillimetre galaxies
evolve into QSOs, with the X-ray absorbed QSOs representing a brief,
transitional phase between these two stages. The discovery that
submillimetre galaxies typically contain more heavily absorbed, lower
luminosity AGN than X-ray absorbed QSOs, supports this evolutionary
picture \citep{alexander05}. In \citet{stevens05} we further proposed
that the X-ray absorbed QSOs are in the process of driving the
interstellar media out of their host galaxies via radiatively driven
winds, in accord with the models developed by \citet{silk98},
\citet{fabian99}, \citet{granato04} and \citet{dimatteo05}.  Now, with
our \xmm\ observations we have shown that the X-ray absorption in
these objects is indeed due to substantial, ionized, QSO-driven
winds. The space density of X-ray absorbed QSOs is $\sim15$ per cent
that of unabsorbed QSOs \citep{page04,silverman05,page06}, implying
that if all QSOs go through an X-ray absorbed phase in their evolution, 
this phase lasts only $\sim15$ per cent as long
as the unabsorbed phase. These relative lifetimes may be a natural
consequence of the high mass outflow rates in the X-ray absorbed QSOs:
for the majority of our X-ray absorbed QSOs we estimate a mass outflow
rate which is $\sim10$ times higher than the accretion rate, implying
that the fuel supply will be depleted 10 times faster in an X-ray
absorbed QSO than in an unabsorbed QSO for the same accretion rate and
fuel reservoir. However, this argument is more complicated if the two types
of QSO fit within the evolutionary sequence discussed above, because
in this case the fuel supply for an unabsorbed QSO will consist only
of the fuel that remains after the X-ray absorbed phase. Nonetheless, if
material is ejected at $\sim$ 10 times the accretion rate for $\sim$ 15 per
cent of the (X-ray absorbed + unabsorbed) QSO lifetime, the total mass of 
material
ejected by the QSO will be of the same order as that accreted. 

The origin of the ionized, absorbing gas is then a fundamental issue for 
the physics of X-ray absorbed QSOs and their evolution. Assuming that 
X-ray absorbed QSOs have the usual AGN structure envisaged in 
geometric unification schemes, the two obvious sources
of material are the accretion disc and the dusty torus. 
If the wind is driven from the
accretion disc, the ejected material must pass through the accretion
disc during the X-ray absorbed phase; in this case it is difficult to
understand why this happens on a timescale which is short compared to the 
lifetime of the QSO. On the other hand, if the ionized wind is driven from the
torus then the timescale for the X-ray absorbed phase corresponds to the time 
required to erode the inner part of the torus and so enlarge substantially its
opening angle; as shown by \citet{krolik01} this timescale can be much shorter
than the lifetime of the QSO. For accretion at the Eddington rate, the 
escape velocities at the distances of the inner edge of the torus (as 
given in Table~\ref{tab:physics}) are between 900 and 1500 km\,s$^{-1}$, 
and scale inversely with the square root of the 
accretion rate in Eddington units. 
Hence, of our X-ray absorbed QSOs, only the BALQSO 
\rxjd\ has
outflow velocities which are too high to plausibly 
be associated with an outflow from the
inner edge of the torus, implying an accretion disc origin for its outflow. 

Viable AGN feedback models which reproduce the $M-\sigma$ relation generally
incorporate the feedback in the form of kinetic energy given to gas 
within the immediate
($\sim$100 pc) environment of the AGN 
\citep[e.g. ][]{fabian99,granato04,dimatteo05,hopkins06,sijacki07}, 
which then interacts with gas on a
larger scale to terminate star formation in the host spheroid. 
It is generally agreed that of order 5 per cent of the AGN 
radiative output must be fed back as kinetic energy in order for the models to
succeed \citep{silk98,wyithe03,dimatteo05}. We can make a simple estimate for
the kinetic energy of the outflows compared to the radiative output of the
X-ray absorbed QSOs as follows. We take the total energy radiated during the X-ray
absorbed phase to be $E_{T}=\epsilon M_{acc} c^{2}$ where $\epsilon$ is the
radiative efficiency, $c$ is the speed of light and
$M_{acc}$ is the mass accreted during the X-ray absorbed phase. The kinetic
energy built up by the outflow is 
$E_{O}=0.5 M_{out} v^{2}$ where $M_{out}$ is the mass supplied
to the outflow during the X-ray absorbed phase and $v$ is the velocity of the
outflow. The fraction of the radiated output of the QSO which is fed back to
the surrounding gas as kinetic energy of the outflow is then
\begin{equation}
\frac{E_{O}}{E_{T}} = \frac{0.5}{\epsilon}\times\frac{M_{out}}{M_{acc}}
\times\frac{v^{2}}{c^{2}}
\label{eq:feedback}
\end{equation}
If we take $M_{out}/M_{acc} = ${\it \.{M}}$_{out}/${\it
  \.{M}}$_{acc}\sim 10$ for the X-ray absorbed phase (as we find for
the majority of our X-ray absorbed QSOs), and a typical outflow
velocity of 8000 km s$^{-1}$ (see Table \ref{tab:uvlines}), we obtain
$E_{O}/E_{T} \sim 4$~per cent. The value of $E_{O}/E_{T}$ so derived
is independent of the radiative efficiency $\epsilon$, because the
$\epsilon$ in Equation \ref{eq:feedback} cancels with that in Equation
\ref{eq:outflow} from which {\it \.{M}}$_{out}/${\it \.{M}}$_{acc}$ is
obtained. On the other hand, the uncertainty on {\it
  \.{M}}$_{out}/${\it \.{M}}$_{acc}$ is in truth very large ($>$ a
factor of 10), depending on the unknown filling factor $f$, and the
square of the outflow velocity $v$, which may not be the same for the
X-ray absorbing phase as the UV absorber.  Nonetheless, $E_{O}/E_{T}
\sim 4$~per cent is quite consistent with the level of feedback
required to terminate star formation and produce the $M-\sigma$
relation, implying that the ionized outflows in X-ray absorbed QSOs
could plausibly be the mechanism through which star formation is
terminated in massive galaxies.

\section{Conclusions}
\label{sec:conclusions}

We have presented X-ray spectra from \xmm\ EPIC and rest-frame
ultraviolet spectra from ground-based telescopes for five X-ray
absorbed, submillimetre-luminous QSOs.  All five QSOs exhibit strong
C\,IV absorption lines in their ultraviolet spectra with equivalent
width $> 5$\AA. The X-ray spectra can be modelled successfully in
terms of $\chi^{2}$ with either cold or ionized absorbers.  The cold
X-ray absorber model requires that the QSOs have unusually flat X-ray
continuum shapes and unusual optical to X-ray spectral energy
distributions, while the ionized absorber model does not require
abnormal underlying continuum properties. This finding, coupled with
the presence of strong C\,IV absorption lines in the UV leads us to
favour the ionized absorber model over the cold absorber
model. Assuming that the X-ray absorbing gas is outflowing with the
same velocities as the C\,IV absorbers, we are able to investigate the
likely location, mass outflow rates and energetics of the ionized
absorbers. We find that the X-ray absorbing gas is likely to be
located within 10\,pc of the continuum source, and so is
associated with the active nucleus rather than the surrounding host
galaxy's interstellar medium.  We estimate that the fraction of
radiated power that is converted into kinetic luminosity of the
outflowing winds is typically $\sim$ 4 per cent, in agreement with
estimates for the kinetic feedback from QSOs that is required to
produce the $M - \sigma$ relation. This finding is thus consistent
with the hypothesis that X-ray absorbed QSOs represent the transition phase between
obscured accretion and the luminous QSO phase in the evolution of
massive galaxies.

\section{Acknowledgments}

Based on observations obtained with \xmm,
an ESA science mission with instruments and contributions directly
funded by ESA Member States and NASA.  This research was also based on
observations made at the William Herschel Telescope which is operated
on the island of La Palma by the Isaac Newton Group in the Spanish
Observatorio del Roque de los Muchachos of the Instituto de
Astrofisica de Canarias, and on observations collected at the European
Southern Observatory, Chile, ESO No. 62.O-0659. We thank the Royal Society 
for travel support under their Joint International Project scheme.  
SRON is supported financially by NWO, the Netherlands Organization for 
Scientific Research.
FJC acknowledges
financial support from the Spanish Ministerio de Ciencia e 
Innovaci\'on (previously Ministerio de Educaci\'on y Ciencia),
under projects ESP2006-13608-C02-01, AYA2009-08059 and AYA2010-21490-C02-01.

\end{document}